\documentclass[epj,twocolumn]{webofc}
\usepackage[varg]{txfonts}   
\woctitle{XLIV International Symposium on Multiparticle Dynamics}
\begin{document}
\title{System size dependence of the log-periodic oscillations of transverse momentum spectra
\thanks{Presented by M.Rybczy\'nski}}
%
%

\author{Maciej Rybczy\'nski\inst{1}\fnsep\thanks{\email{Maciej.Rybczynski@ujk.edu.pl}} \and
        Grzegorz Wilk\inst{2}\fnsep\thanks{\email{Grzegorz.Wilk@fuw.edu.pl}} \and
        Zbigniew W\l odarczyk\inst{1}\fnsep\thanks{\email{Zbigniew.Wlodarczyk@ujk.edu.pl}}
}

\institute{Institute of Physics, Jan Kochanowski University,
Kielce, Poland \and
           National Centre for Nuclear Research, Warsaw, Poland
}

\abstract{ Recently the inclusive transverse momentum
distributions of primary charged particles were measured for
different centralities in $Pb+Pb$ collisions. A strong suppression
of the nuclear modification factor in central collisions around
$p_T \sim 6-7$~GeV/c was seen. As a possible explanation, the
hydrodynamic description of the collision process was tentatively
proposed. However, such effect, (albeit much weaker) also exists
in the ratio of data/fits, both in nuclear $Pb+Pb$ collisions, and
in the elementary $p+p$ data in the same range of transverse
momenta for which such an explanation is doubtful. As shown
recently, in this case, assuming that this effect is genuine, it
can be attributed to a specific modification of a quasi-power like
formula usually used to describe such $p_T$ data, namely the
Tsallis distribution. Following examples from other branches of
physics, one simply has to allow for the power index becoming a
complex number. This results in specific log-periodic oscillations
dressing the usual power-like distribution, which can fit the
$p+p$ data. In this presentation we demonstrate that this method
can also describe $Pb+Pb$ data for different centralities. We
compare it also with a two component statistical model with two
Tsallis distributions recently proposed showing that data at still
larger $p_T$ will be sufficient to discriminate between these two
approaches.}
\maketitle
\section{Introduction}
\label{Sec:Intro}

Recently the inclusive transverse momentum distributions of
primary charged particles were measured for different centralities
in $Pb+Pb$ collisions at $\sqrt{s_{NN}}=2.76$~TeV~
\cite{CMS:2012aa,Abelev:2012hxa}. Data were presented in terms of
a nuclear modification factor which exhibits strong suppression in
the case of central collisions for $p_T$ around $6-7$~GeV/c
\cite{CMS:2012aa,Abelev:2012hxa}. For more peripheral collisions
this suppression becomes weaker but never vanishes. Several
theoretical models based on a hydrodynamic description of the
medium were proposed to study and explain the effect
\cite{CMS:2012aa,Abelev:2012hxa}. Because, for some time already,
it became popular to fit the different kinds of transverse
momentum spectra measured in multiparticle production processes
using a statistical approach based on a nonextensive quasi-power
Tsallis formula (cf. Eq.~(\ref{Eq:f}) below)
\cite{TsallisUseWW,TsallisUseCW,TsallisUseBi,TsallisUseDe,TsallisUseRW,TsallisUseOt},
such an approach was also used. However, it turned out that one
needs at least a two component statistical model with two Tsallis
distributions \cite{Biro:2014cka,Urmossy:2014gpa} (in these papers
these two components were identified with, respectively, {\it
soft} and {\it hard} dynamics of the underlining production
process) \footnote{Similar idea of multicomponent approach, but
based on a revised two- or multi-Boltzmann distribution instead of
Tsallis distribution, was presented also in \cite{BBB} (albeit it
was applied there only to a rather limited range of transverse
momenta). }.

In the mean time it was realized that, when looking at the ratio
{\it data/fit} for data on the $p_T$ distributions from $p+p$
collisions, taken in the same range of transverse momenta and at
the same energies as data from $Pb+Pb$ collisions
\cite{CMS,ALICE,ATLAS} (and which, as shown in~\cite{CYWW}, can be
adequately fitted by a single Tsallis formula), one then observes
strong modulation in $Pb+Pb$ data and a similar (albeit much
weaker) modulation also in $p+p$ data. However, for $p+p$ data any
explanation based on hydrodynamic models is doubtful. Therefore
the natural question one can pose is whether a single nonextensive
Tsallis formula so successful in fitting $p_T$ data
\cite{TsallisUseWW,TsallisUseCW,TsallisUseBi,TsallisUseDe,TsallisUseRW,TsallisUseOt,CYWW},
can also describe these results. As shown in
\cite{WW_LPO,Wilk:2014bia} the answer is positive, here we extend
the analysis presented there for the $p+p$ collisions to analysis
of dependence of the shape of the $p_T$ spectra of charged hadrons
produced in $Pb+Pb$ collisions and its dependence on the size of a
colliding system.

Transverse momentum distributions measured in $p+p$ interactions
exhibit for large $p_T$ roughly a power-like behavior, whereas
they become purely exponential for small $p_T$. For a long time
already, for different reasons, it was found reasonable to use
instead of two different formulas for these two parts of phase
space (reflecting, as it is believed, different dynamics operating
there), some single interpolating formula \cite{CM}. The most know
at present is its version known as {\it QCD inspired Hagedorn
form} \cite{H} (with parameters: $m$ and $T$)
\begin{equation}
h\left(p_{T}\right)=C\cdot\left(1+\frac{p_{T}}{m\cdot
T}\right)^{-m}. \label{Eq:h}
\end{equation}
The other is  a Tsallis formula \cite{T} (with two parameters: $q$
and $T$, $C$ is a normalization constant)
\begin{equation}
f\left(p_{T}\right)=C\cdot\left[1-\left(1-q\right)\frac{p_{T}}{T}\right]^{1/(1-q)}.
\label{Eq:f}
\end{equation}
For our purposes, both formulas are equivalent for $m=1/(q-1)$
(and we shall use them interchangeably.) They both represent the
simplest way of describing the whole observed range of measured
$p_{T}$ distributions (and they both replace $p_0$, a momentum
customarily used to separate {\it soft} and {\it hard} parts of
the momentum phase space, by the parameter $T$, which can be
interpreted either as a scale parameter for the hard component or
a kind of temperature for the soft one). The best examples are the
recent successful fits \cite{CYWW} to very large $p_{T}$ data
measured by the LHC experiments CMS~\cite{CMS}, ATLAS~\cite{ATLAS}
and ALICE~\cite{ALICE} for $p+p$ collisions. However, albeit these
fits look pretty good, closer inspection shows that the ratio of
{\it data/fit} is not flat but shows some kind of specific
log-periodic oscillations as a function of $p_T$
\cite{WW_LPO,Wilk:2014bia} (cf. also \cite{CTWW,GW}).

At this point one should remember that in other branches of
science, whenever one encounters pure power-like distributions,
one finds that in many cases these distributions are {\it
decorated} by specific log-periodic oscillations, i.e., they are
multiplied by some {\it dressing factor} $R$, which is customarily
taken in the form of \cite{LPO-examples}:
\begin{equation}
R\left(p_{T}\right)=a+b\cos\left[c\cdot\ln\left(p_{T}+d\right)+f\right].
\label{Eq:R}
\end{equation}
(here presented as function of $p_T$, which is our variable).

\section{Derivation of the dressing factor $R$ for Tsallis distributions}
\label{Sec:Scale_inv}

Log-periodic oscillations are usually regarded as an indication of
the presence of some hierarchical, multiscale, fine-structure,
most probably of some kind of (multi) fractal origin. In what
concerns oscillations apparently seen in \cite{CMS,ALICE,ATLAS}
data, which is our case,  it was assumed in
\cite{WW_LPO,Wilk:2014bia} that they are not an experimental
artifact but, rather, that they are caused by some genuine
dynamical effect and, as such, they should be studied carefully.
The rationale was that these oscillations are seen by all
experiments, at all energies at which data were taken, and show
almost identical patterns. Now, in addition, they are changing
with size of the colliding objects, as seen in the $Pb+Pb$ data.
As in other places, where such oscillations were investigated for
simple power-like distributions \cite{LPO-examples}, we further
assumed that to account for them the original Tsallis formula
(either $h\left(p_{T}\right)$) from Eq.~(\ref{Eq:h}) or
$f\left(p_{T}\right)$ from Eq.~(\ref{Eq:f})) has to be multiplied
by a log-oscillating function $R$, as given by Eq.~(\ref{Eq:R}),
with the parameters connected to the original parameters of the
respective form of Tsallis distribution used. For completeness of
the presentation we shall repeat shortly its derivation (cf.
\cite{WW_LPO,Wilk:2014bia} for details).

Start from the simple pure power law distribution,
\begin{equation}
O\left(x\right)=C\cdot x^{-m} \label{Eq:power}
\end{equation}
This function is scale invariant, i.e.,
\begin{equation}
O\left(\lambda x\right)=\mu O\left( x\right) \label{Eq:sinv}
\end{equation}
with $m=-\ln{\mu}/\ln{\lambda}$. However, because $1 =
\exp{\left(\imath 2\pi k\right)}$,
\begin{equation}
\mu\lambda^{m} = 1 = \exp{\left(\imath 2\pi k\right)},\quad k = 0,
1, \dots , \label{Eq:kkk}
\end{equation}
i.e., it means that, in general,  the index $m$ can become
complex,
\begin{equation}
m=-\frac{\ln\mu}{\ln\lambda}+\imath\frac{2\pi k}{\ln\lambda}.
\end{equation}
Such form of the power index results in $R$ as given by Eq.~(\ref{Eq:R}) when one keeps only $k = 0, 1$ terms~\cite{LPO-examples}.

However, Tsallis distribution is not a pure power-law but rather a
quasi-power distribution, it contains a scale $T$ for the variable
considered and it has also a constant term, as in
Eqs.~(\ref{Eq:h}) or (\ref{Eq:f}). One must therefore find a
variable in which our Tsallis distribution will show scaling
property. It turns out that the evolution of the differential
$df\left(p_{T}\right)/dp_{T}$ of a Tsallis distribution
$f\left(p_{T}\right)$ with power index $n$ performed for finite
difference, $\delta p_{T}=\alpha\left(nT+p_{T}\right)$, replacing
differential $d p_T$ (here $\alpha$ is a kind of scaling factor,
for $\alpha \rightarrow 0$ oscillations vanish) and using variable
\begin{equation}
x = 1 + \frac{p_T}{nT} \label{Eq:varX}
\end{equation}
results in the desired scale invariant relation, which in our case
takes the form of (cf. \cite{WW_LPO,Wilk:2014bia} for details of
derivation):
\begin{equation}
g\left[\left(1+\alpha\right)x\right]=\left(1-\alpha
n\right)g\left(x\right),
\end{equation}
This means that one can write Eq.~(\ref{Eq:h}) in the form:
\begin{equation}
g\left(x\right)=x^{-m_{k}}, \label{Eq:g}
\end{equation}
with
\begin{equation}
m_{k}=-\frac{\ln \left(1-\alpha n\right)}{\ln
\left(1+\alpha\right)}+\imath k\frac{2\pi}{\ln
\left(1+\alpha\right)}, \label{Eq:MM}
\end{equation}
or, more generally, as the sum
\begin{eqnarray}
g\left(x\right) &=& \sum_{k=0}^{\infty}w_{k}\cdot
Re\left(x^{-m_{k}}\right)=\nonumber\\
&=& x^{-Re\left(m_{k}\right)}\sum_{k=0}^{\infty}w_{k}\cdot\cos\left[Im\left(m_{k}\right)\ln
x\right]. \label{Eq:Tg}
\end{eqnarray}
Since we do not know {\it a priori} the details of the dynamics of
processes under consideration (i.e. we do not know the weights
$w_{k}$), in what follows we use (as before) only the two first
terms, $k=0$  and $k=1$, getting dressed Tsallis distribution
\begin{equation}
g\left(p_{T}\right) = \left(1+\frac{p_{T}}{nT}\right)^{m_{0}}\cdot
R\left(p_T\right) \label{Eq:gTsallis}
\end{equation}
with dressing factor which now has the following form:
\begin{equation}
R\left( p_T\right) \simeq \left\{w_{0}+w_{1}\cos\left[\frac{2\pi}
{\ln\left(1+\alpha\right)}\ln\left(1+\frac{p_{T}}
{nT}\right)\right]\right\}. \label{Eq:TsallisR}
\end{equation}
The parameters in general modulating factor given by
Eq.~(\ref{Eq:R}) are now identified as follows:
\begin{eqnarray}
&& a=w_{0},\quad b=w_{1},\quad
c=2\pi/\ln\left(1+\alpha\right),\nonumber\\
&& d = nT,\quad f=-c\cdot\ln\left(nT\right.) \label{eq:parameters}
\end{eqnarray}
Notice that this {\it dressing procedure} introduces three new
parameters: scaling factor $\alpha$ and weights $w_0$ and $w_1$.

\section{Transverse momentum distributions in Pb+Pb collisions}
\label{Sec:Tr_mom_dist}

\begin{figure}[h]
\centering
\includegraphics[scale=0.34]{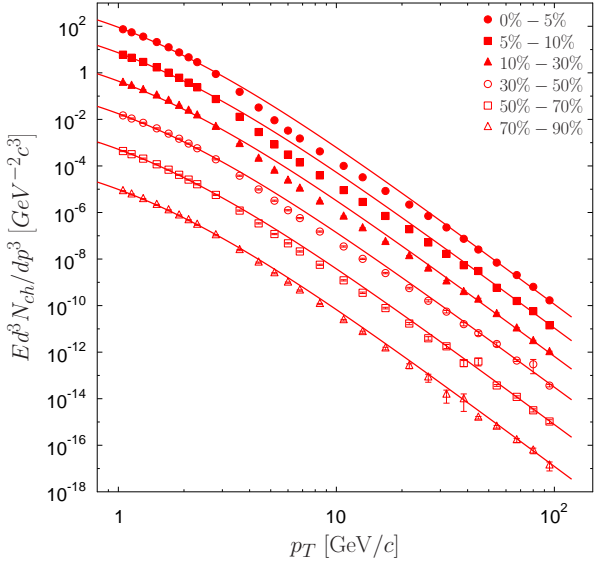}
\caption{Transverse momentum distributions of particles produced
in Pb+Pb collisions at $\sqrt{s_{NN}}=2.76$~TeV~\cite{ALICE} fitted by
Eq.~(\ref{Eq:h}). For better readability, results for different
centralities are scaled by $10^{-i}$, $i=0, 1, 2, \cdots, 5$ from
the most central to the most peripheral collisions.}
\label{Fig:pt_fits}
\end{figure}

\begin{figure}[h]
\centering
\includegraphics[scale=0.34]{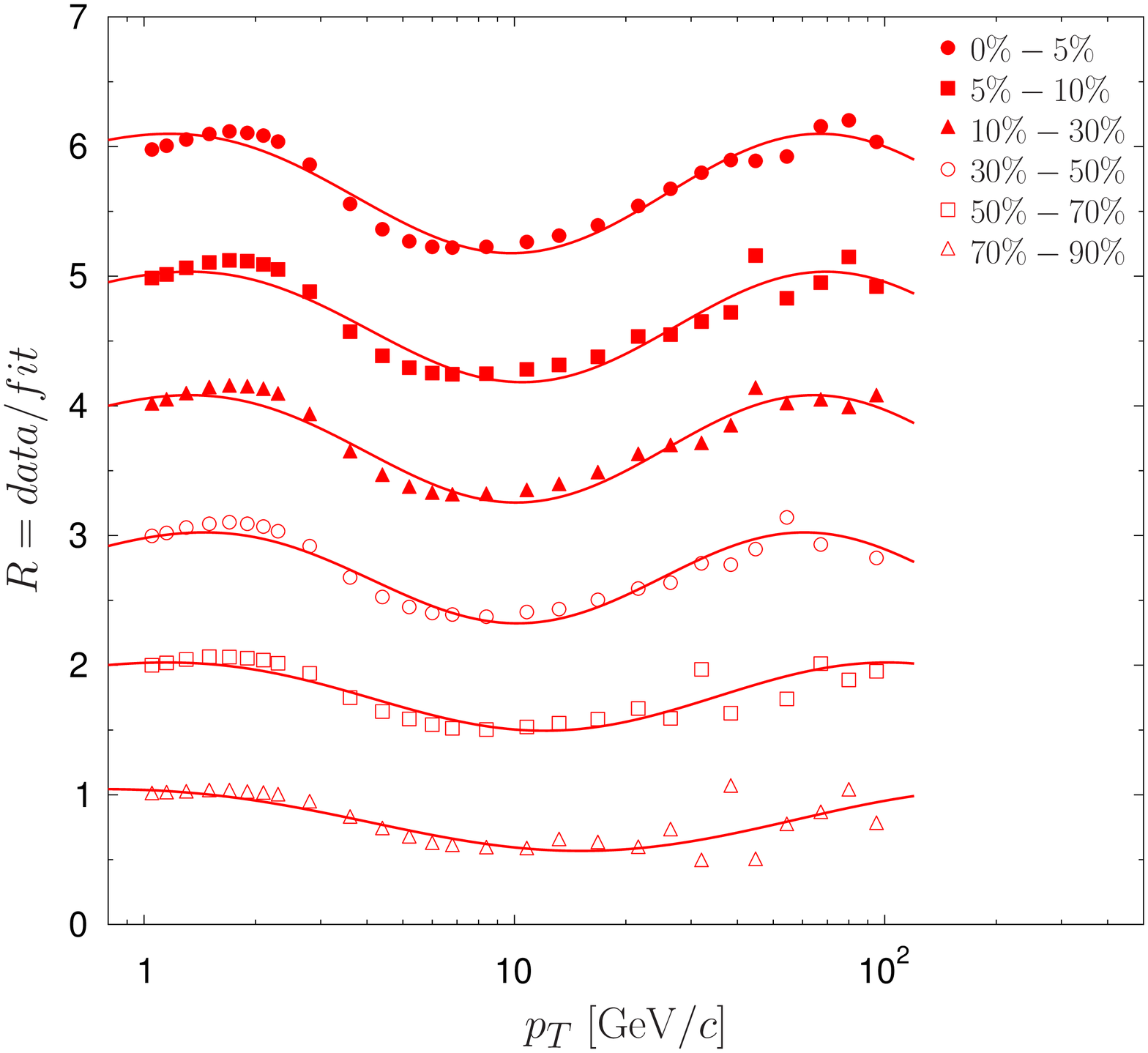}
\caption{Data/fit ratio from the Fig.~\ref{Fig:pt_fits} fitted by
Eq.~(\ref{Eq:R}). For better readability, results for different
centralities are shifted by $i=0, 1, 2, \cdots, 5$ from the most
peripheral to the most central collisions.} \label{Fig:pt_ratios}
\end{figure}

Recently, the inclusive transverse momentum distributions of
primary charged particles are measured for different centralities
in $Pb+Pb$ collisions at $\sqrt{s_{NN}}=2.76$
TeV~\cite{CMS:2012aa,Abelev:2012hxa}, see Fig.~\ref{Fig:pt_fits}.
The data, presented in terms of nuclear modification factor, show
strong suppression in central collisions for $p_T$ around $6-7$
GeV/c \cite{CMS:2012aa,Abelev:2012hxa}. As shown in
Fig.~\ref{Fig:pt_fits}, these data can be fitted using a Tsallis
distribution in the form of Eq.~(\ref{Eq:h}) with parameters as
listed in Table~\ref{Tab:pt_fits}. In Fig.~\ref{Fig:pt_ratios} we
show {\it data/fit} ratios, which exhibit rather dramatic
log-oscillatory structure, increasing for most central collisions.
As shown there it can be fitted, for all centralities, by using a
dressing factor $R$ as defined in Eq.~(\ref{Eq:R}), with
parameters listed in Table~\ref{Tab:pt_ratios}.
\begin{table}[h]
\centering \caption{Parameters used in Eq.~(\ref{Eq:h}) to fit the
spectra presented in Figs.~\ref{Fig:pt_fits}-\ref{Fig:70-90_PP}}
\label{Tab:pt_fits}
\begin{tabular}{cccc}
\hline $centrality~[\%]$ & $C$ & $m$ & $T$  \\\hline
$0-5$ & $11000$ & $7.0$ & $0.145$ \\
$5-10$ & $8750$ & $6.95$ & $0.145$ \\
$10-30$ & $5300$ & $6.95$ & $0.145$ \\
$30-50$ & $2100$ & $6.9$ & $0.145$ \\
$50-70$ & $625$  & $6.95$ & $0.145$ \\
$70-90$ & $119$ & $7.0$ & $0.145$ \\\hline $p+p$ & $22.65$ & $7.1$
& $0.145$\\\hline
\end{tabular}
\end{table}

\begin{figure}[h]
\centering
\includegraphics[scale=0.34]{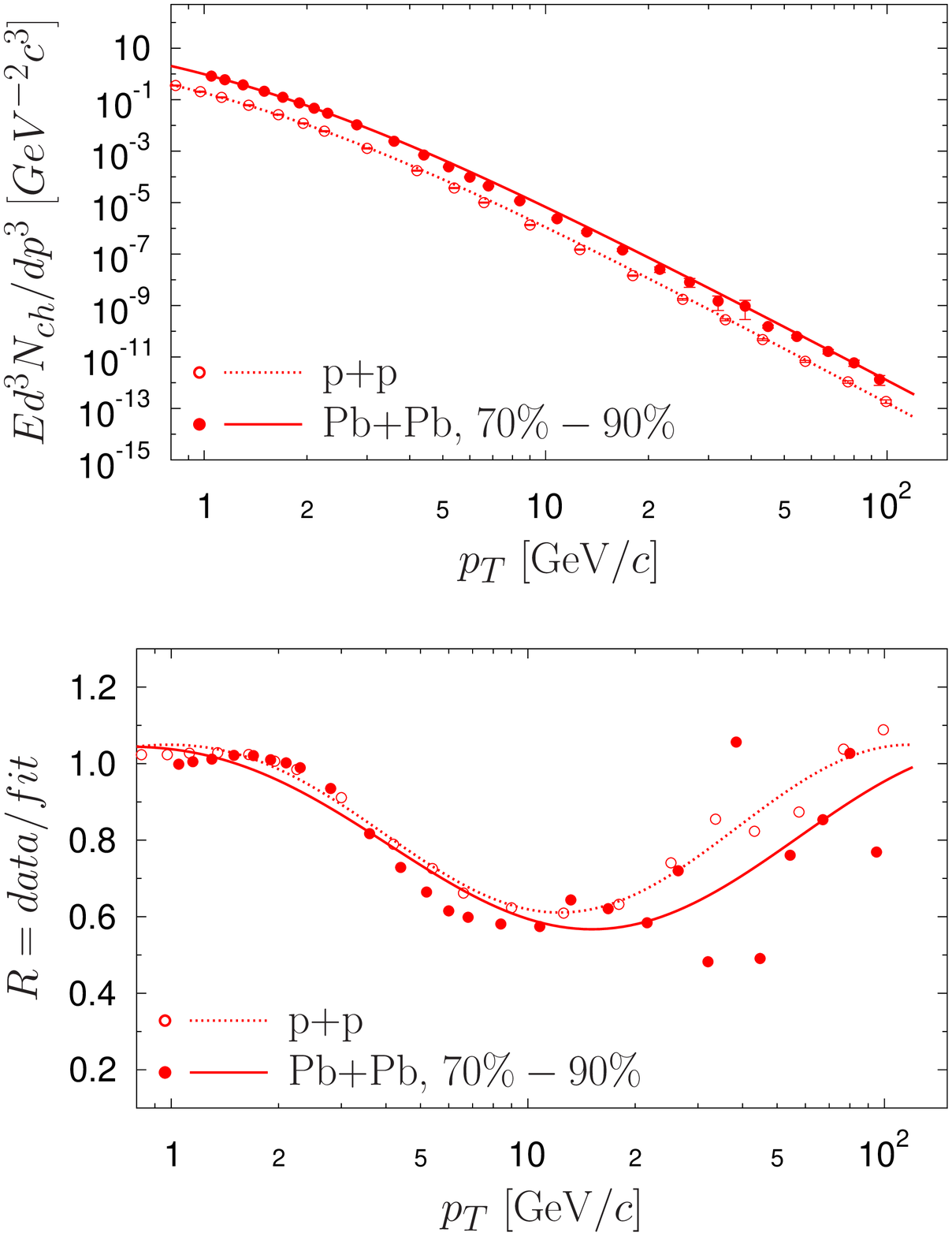}
\caption{Comparison of results for $p_T$ distributions in $p+p$
collisions with the most peripheral $Pb+Pb$ ones. Data taken from~\cite{ALICE}.}
\label{Fig:70-90_PP}
\end{figure}

From Fig.~\ref{Fig:pt_ratios} one can deduce that the amplitude of
the oscillating term in Eq.~(\ref{Eq:R}) reaches its maximum for
the most central collisions, and smoothly decreases when going to
more peripheral interactions. As seen in Fig.~\ref{Fig:70-90_PP}
there is also reasonable agreement between results from $p+p$ and
most peripheral $Pb+Pb$ collisions. However, closer look at
parameters reveal that parameters $c$ and $f$ differs
substantially in both cases (cf. Table \ref{Tab:pt_ratios}). This
can be attributed to changes in the scaling parameter $\alpha$ in
Eq.~(\ref{eq:parameters}).

\begin{table*}[t]
\centering \caption{Parameters used in Eq.~(\ref{Eq:R}) to fit the
data/fit ratios presented in
Figs.~\ref{Fig:pt_ratios}-\ref{Fig:70-90_PP}}
\label{Tab:pt_ratios}
\begin{tabular}{cccccc}
\hline $centrality~[\%]$ & $a$ & $b$ & $c$ & $d$ & $f$ \\\hline
$0-5$ & $0.638$ & $0.461$ & $1.664$ & $0.368$ & $-0.719$ \\
$5-10$ & $0.609$ & $0.426$ & $1.690$ & $0.368$ & $-0.889$ \\
$10-30$ & $0.668$ & $0.414$ & $1.725$ & $0.368$ & $-0.910$ \\
$30-50$ & $0.673$ & $0.351$ & $1.789$ & $0.368$ & $-1.072$ \\
$50-70$ & $0.757$  & $0.263$ & $1.494$ & $0.368$ & $-0.622$ \\
$70-90$ & $0.806$ & $0.238$ & $1.199$ & $0.368$ & $-0.143$
\\\hline $p+p$ & $0.830$ & $0.219$ & $1.407$ & $0.368$ & $-0.428$
\\\hline
\end{tabular}
\end{table*}

\subsection{Centrality dependence of parameters}

\begin{figure}[h]
\centering
\includegraphics[scale=0.39]{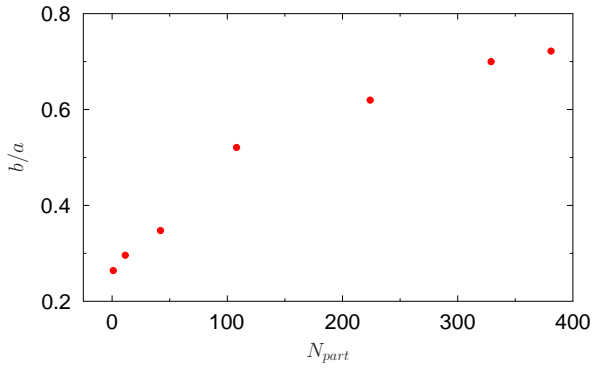}
\caption{Ratio of the values of $b/a$ parameters obtained from the
fits presented on Fig.~\ref{Fig:pt_ratios} as a function of number
of nucleons participating in collision, $N_{part}$.}
\label{Fig:par1}
\end{figure}

When analyzing results from nuclear collisions it is customarily
to look especially for their sensitivity on the number of nucleons
participating in collision (i.e., participants, $N_{part}$) and on
the number of binary nucleon-nucleon collisions, $N_{coll}$. On
Fig.~\ref{Fig:par1} the ratio of oscillating to constant term,
$b/a$ in Eq.~(\ref{Eq:R}), is presented as a function of
$N_{part}$. One observes a smooth increase of this ratio when
going to more central collisions.
\begin{figure}[h]
\centering
\includegraphics[scale=0.39]{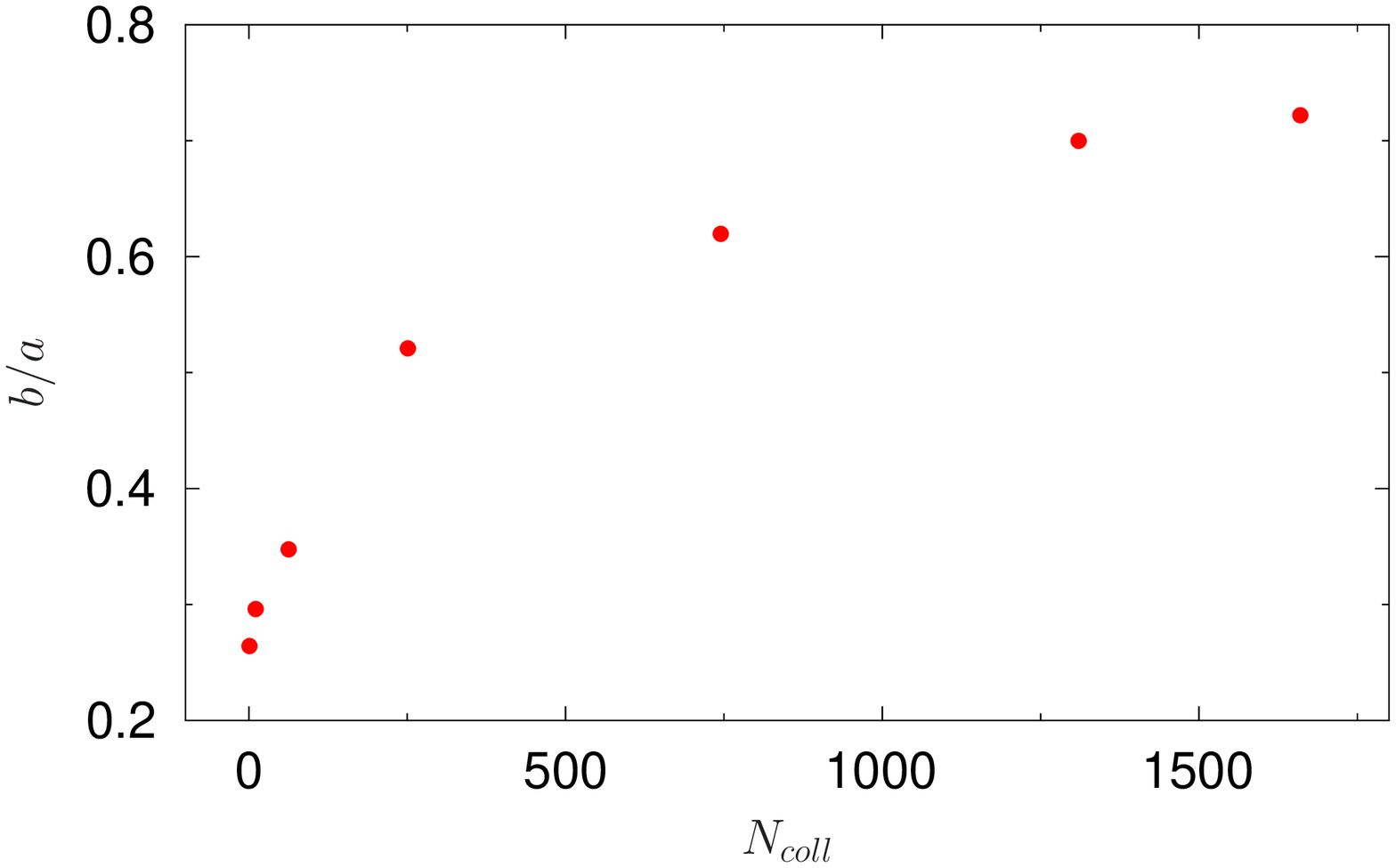}
\caption{Ratio of the values of $b/a$ parameters obtained from the
fits presented on Fig.~\ref{Fig:pt_ratios} as a function of number
of collisions, $N_{coll}$.} \label{Fig:par2}
\end{figure}
\begin{figure}[h]
\centering
\includegraphics[scale=0.39]{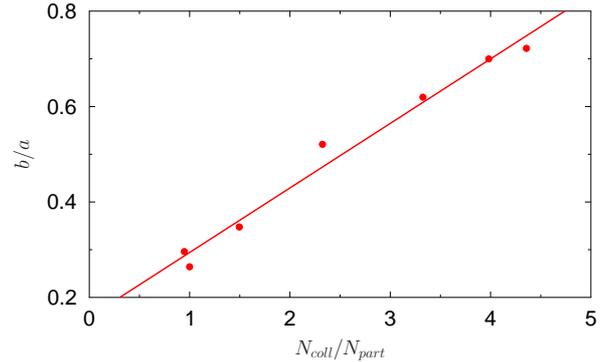}
\caption{Ratio of the values of $b/a$ parameters obtained from the
fits presented on Fig.~\ref{Fig:pt_ratios} as a function of number
of collisions per participant nucleon, $N_{coll}/N_{part}$.}
\label{Fig:par3}
\end{figure}
On Fig.~\ref{Fig:par2} the same ratio is plotted as a functions of
$N_{coll}$. Again, there is a smooth increase of the $b/a$ ratio
with increasing number of collisions. Interestingly, when plotting
the ratio $b/a$ as a function  of the number of collisions per the
number of participants, $N_{coll}/N_{part}$, one finds a visible
linear increase, see Fig.~\ref{Fig:par3}. Such behavior suggests
that the influence of oscillatory part increases with increasing
percentage of binary collisions (usually attributed to hard
scattering, possibly to particles produced from jets).

\subsection{Two component Tsallis fit}

\begin{figure}[h]
\centering
\includegraphics[scale=0.40]{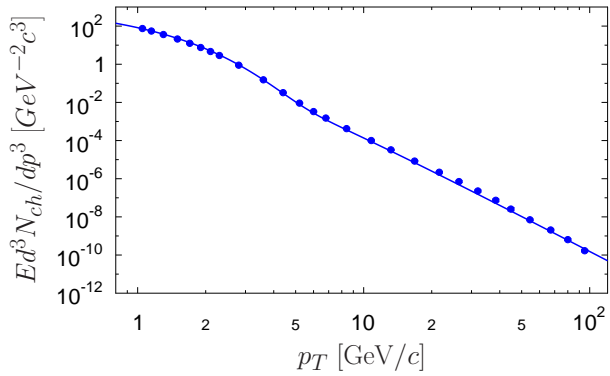}
\caption{Transverse momentum distribution of particles produced in
most central, $c=0-5\%$, Pb+Pb collisions at
$\sqrt{s_{NN}}=2.76$~TeV~\cite{ALICE} fitted by Eq.~(\ref{Eq:h2}) with
parameters: $C_{1}=419.133$, $C_{2}=1100.0$, $T_{1}=0.158$, $T_{2}=0.369$, $m_{1}=6.172$, and $m_{2}=66.660$.} \label{Fig:two_tsallis}
\end{figure}

As mentioned already before, recently a different method has been
proposed to describe a structure clearly visible in the transverse
momentum spectra obtained in $Pb+Pb$ collisions
\cite{Biro:2014cka,Urmossy:2014gpa} (see also \cite{BBB}).
Recognizing that a single Tsallis fit is not able to fully
reproduce the observed structure, it was argued that one should
resort to two power-laws, i.e., to two Tsallis distributions.
According to these authors, they could be attributed to two
possible mechanisms of particle production, also mentioned before,
soft and hard, each with different sensitivity to $N_{part}$. The
increased (as in our case) number of parameters allows them to
obtain very good fits, cf., Fig.~\ref{Fig:two_tsallis}, where we
present an example of the use of this method for the most central,
$c=0-5\%$, Pb+Pb collisions at $\sqrt{s_{NN}} = 2.76$~TeV using
the following double-Tsallis formula:
\begin{equation}
h_{2}\left(p_{T}\right)=C_{1}\cdot\left(1+\frac{p_{T}}{m_{1}
T_{1}}\right)^{-m_{1}}+ C_{2}\cdot\left(1+\frac{p_{T}}{m_{2}
T_{2}}\right)^{-m_{2}} , \label{Eq:h2}
\end{equation}
\begin{table}[h]
\centering \caption{Parameters used in Eq.~(\ref{Eq:h}) to fit the
spectra presented } \label{Tab:LCfits}
\begin{tabular}{cccc}
\hline $centrality~[\%]$ & $C$ & $m$ & $T$  \\\hline
$0-5$ & $11000$ & $7.0$ & $0.145$ \\
$5-10$ & $8750$ & $6.95$ & $0.145$ \\
$10-30$ & $5300$ & $6.95$ & $0.145$ \\
$30-50$ & $2100$ & $6.9$ & $0.145$ \\
$50-70$ & $625$  & $6.95$ & $0.145$ \\
$70-90$ & $119$ & $7.0$ & $0.145$ \\\hline $p+p$ & $22.65$ & $7.1$
& $0.145$\\\hline
\end{tabular}
\end{table}

\begin{figure}
\centering
\includegraphics[scale=0.37]{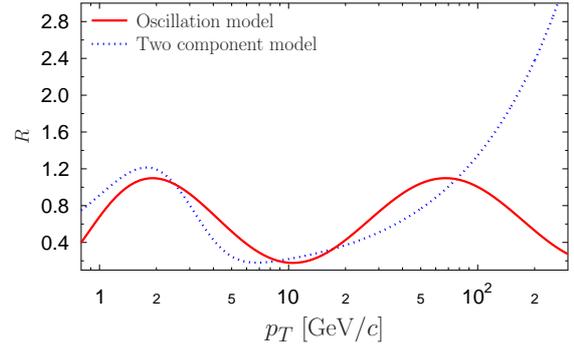}
\caption{Ratios of two component fit to single Tsallis fit (dotted
line), and Tsallis with oscillating term (Eq.~(\ref{Eq:R})) for
oscillation model (solid line) plotted for the most central,
$c=0-5\%$, Pb+Pb collisions at $\sqrt{s_{NN}}=2.76$~TeV.}
\label{Fig:ratios}
\end{figure}

The result presented on Fig. \ref{Fig:two_tsallis} is indeed
encouraging. Regarding Eq.~(\ref{Eq:h2}) as effectively the two
first terms in a kind of expansion of our dressed Tsallis
formula~(\ref{Eq:gTsallis}), it is not surprising that Eq.
(\ref{Eq:h2}) vaguely follows our results, as can be seen in Fig.
\ref{Fig:ratios}. It presents the comparison of the ratio of best
fits obtaining from both formulas, respectively Eq.~(\ref{Eq:h2})
and Tsallis formula with oscillating term, Eq.~(\ref{Eq:R}), to a
single Tsallis fit, Eq.~(\ref{Eq:h}), plotted for the most
central, $c=0-5\%$, Pb+Pb collisions at $\sqrt{s_{NN}}=2.76$~TeV.

However, both approaches diverge dramatically for large values of
$p_T$. This is because the log-oscillating formula
(\ref{Eq:TsallisR}) contains term which can be positive or
negative whereas the two-Tsallis one is always positive and
therefore the ratio plotted in Fig. \ref{Fig:two_tsallis} must
ultimately grow as a function of $p_T$. Should the data in that
region be available, it would allow us to decide which model
better describes the mechanism of particle production in
relativistic heavy ion collisions.

\section{Summary}
\label{Sec:Summ}

Recently, the inclusive transverse momentum distributions of
primary charged particles were measured for different centralities
in $Pb+Pb$ collisions at $\sqrt{s_{NN}}=2.76$~TeV. Data presented
in terms of the nuclear modification factor show a strong
suppression in central collisions for $p_{T}$ around $6-7$~GeV/c.
The dependence of the shape of the $p_{T}$~spectra on the size of
a colliding system has been discussed using Tsallis distribution
as a reference spectrum. It is remarkable that this kind of
suppression is also observed in $p+p$ collisions where the
mechanism of particle production is believed to be different. This
suppression, visualized as possibly signal of log-oscillatory
behavior known from other branches of physics, has been attributed
there to imaginary part in the Tsallis power index. When analyzing
$Pb+Pb$ data along the same lines, one gets that the amplitude of
the corresponding (much stronger than in $p+p$ case) oscillations
increases linearly as a function of number of collisions per
participant nucleon, $N_{coll}/N_{part}$. We compared our results
with recent proposition of using two-power laws Tsallis fits to
describe such data and proposed way of experimental
differentiating between these approaches.\\

Acknowledgments: This research  was supported in part by the
National Science Center (NCN) under contract Nr
2013/08/M/ST2/00598. We would like to warmly thank Dr Eryk Infeld
for reading this manuscript.\\

\end{document}